\documentclass[
twocolumn,
english,
aps,
pra,
longbibliography,
superscriptaddress,
amsmath,
amssymb,
floatfix,
]{revtex4-2}
\usepackage[utf8]{inputenc}
\usepackage{microtype}
\usepackage{graphicx}
\usepackage{amsmath}
\usepackage{amssymb}
\usepackage{mathtools}
\usepackage[colorlinks=true,urlcolor=blue,citecolor=blue,linkcolor=blue]{hyperref}
\usepackage{natbib}
\usepackage{physics}
\usepackage{cleveref}
\usepackage{sidecap}
\usepackage{makecell}
\usepackage{listings}
\usepackage[linesnumbered,lined,commentsnumbered]{algorithm2e}
\Crefname{algocf}{Algorithm}{Algorithms}
\usepackage{layouts}

\begin{document}

\title{Combining quantum processors with real-time classical communication}
\date{\today}
\author{Almudena Carrera Vazquez}
\affiliation{IBM Quantum, IBM Research Europe - Zurich, R\"uschlikon 8803, Switzerland}
\author{Caroline Tornow}
\affiliation{IBM Quantum, IBM Research Europe - Zurich, R\"uschlikon 8803, Switzerland}
\affiliation{Institute for Theoretical Physics, ETH Zurich 8093, Switzerland}
\author{Diego Rist\`e}
\affiliation{IBM Quantum, IBM Research Cambridge, Cambridge, MA 02142, USA}
\author{Stefan Woerner}
\affiliation{IBM Quantum, IBM Research Europe - Zurich, R\"uschlikon 8803, Switzerland}
\author{Maika Takita}
\affiliation{IBM Quantum, T. J. Watson Research Center, Yorktown Heights, NY, USA}
\author{Daniel J. Egger}
\email{deg@zurich.ibm.com}
\affiliation{IBM Quantum, IBM Research Europe - Zurich, R\"uschlikon 8803, Switzerland}

\begin{abstract}
Quantum computers process information with the laws of quantum mechanics. 
Current quantum hardware is noisy, can only store information for a short time, and is limited to a few quantum bits, i.e., qubits, typically arranged in a planar connectivity. 
However, many applications of quantum computing require more connectivity than the planar lattice offered by the hardware on more qubits than is available on a single quantum processing unit (QPU). 
Here we overcome these limitations with error mitigated dynamic circuits and circuit-cutting to create quantum states requiring a periodic connectivity employing up to 142 qubits spanning multiple QPUs connected in real-time with a classical link. 
In a dynamic circuit, quantum gates can be classically controlled by the outcomes of mid-circuit measurements within run-time, i.e., within a fraction of the coherence time of the qubits. 
Our real-time classical link allows us to apply a quantum gate on one QPU conditioned on the outcome of a measurement on another QPU which enables a modular scaling of quantum hardware.
Furthermore, the error mitigated control-flow enhances qubit connectivity and the instruction set of the hardware thus increasing the versatility of our quantum computers. 
Dynamic circuits and quantum modularity are thus key to scale quantum computers and make them useful.
\end{abstract}

\maketitle{}

\section{Introduction}

Quantum computers promise to deeply impact a wide variety of domains ranging from natural sciences~\cite{Dimeglio2023, Bauer2020}, to optimization~\cite{Abbas2023}, and finance~\cite{Egger2020}.
These machines process information encoded in quantum bits with unitary operations.
However, quantum computers are noisy and most large-scale architectures arrange the physical qubits in a planar lattice.
For instance, superconducting qubits are typically arranged in a two-dimensional grid or a heavy-hexagonal layout~\cite{Krantz2019}.
Nevertheless, current processors with error mitigation can already simulate hardware-native Ising models with 127 qubits and measure observables at a scale where classical computers begin to struggle~\cite{Kim2023}.
Crucially, the usefulness of quantum computers hinges on further scaling and overcoming their limited qubit connectivity.
A modular approach is key to scale current noisy quantum processors~\cite{Bravyi2022} and to achieve the large numbers of physical qubits needed for fault tolerance~\cite{Bravyi2023}.
In the near term, modularity in superconducting qubits is achieved by short-range inter-connects which link adjacent chips~\cite{Conner2021, Gold2021}. 
In the medium term, long-range gates operating in the microwave regime may be carried out over long conventional cables~\cite{Zhong2019, Zhong2021, Malekakhlagh2024}.
This would enable a non-planar qubit connectivity suitable for an efficient error correction~\cite{Bravyi2023}.
A long term alternative is to entangle remote QPUs with an optical link leveraging a microwave to optical transduction~\cite{Ang2022}, a feat that has not yet been demonstrated.
In addition, dynamic circuits broaden the set of operations of a quantum computer by performing mid-circuit measurements (MCM) and classically controlling a gate within the coherence time of the qubits.
They enhance algorithmic quality~\cite{Corcoles2021} and qubit connectivity~\cite{Baumer2023}.
As we will show, dynamic circuits also enable modularity by connecting QPUs in real time through a classical link.

We take a complementary approach based on virtual gates and dynamic circuits to implement long-range gates in a modular architecture.
We connect qubits at arbitrary locations with a real-time classical connection and create the statistics of entanglement through a quasi-probability decomposition (QPD)~\cite{Hofmann2009, Piveteau2023}.
More specifically, we consume virtual Bell pairs in a teleportation circuit to implement two-qubit gates~\cite{Gottesman1999, Wan2019}.
On quantum hardware with a sparse and planar connectivity creating a Bell pair between arbitrary qubits requires a long-range controlled-NOT (CNOT) gate.
To avoid these gates we employ a QPD over local operations resulting in \emph{cut Bell pairs} which the teleportation consumes.
We compare the implementation of this Local Operations and Classical Communication (LOCC) scheme to one based on Local Operations (LO) only~\cite{Mitarai2021}.
LO do not need the classical-link and are thus simpler to implement than LOCC.
However, since LOCC only requires a single parameterized template circuit it is more efficient to compile than LO and enables measurement-based quantum computing~\cite{Baumer2023}.

Our work makes four key contributions.
First, Sec.~\ref{sec:cutting} presents the quantum circuits and QPD to create multiple cut Bell pairs to realize the virtual gates of Ref.~\cite{Piveteau2023}.
Second, we suppress and mitigate the errors arising from the latency of the classical control hardware in dynamic circuits~\cite{Viola1999} with a combination of dynamical decoupling and zero-noise extrapolation~\cite{Temme2017}.
Third, in Sec.~\ref{sec:103nodes} we leverage these methods to engineer periodic boundary conditions on a 103 node graph state.
Fourth, in Sec.~\ref{sec:134_nodes} we demonstrate a real-time classical connection between two separate QPUs.
Combined with dynamic circuits, this enables us to operate both chips as a single quantum computer, which we exemplify by engineering a periodic graph state that spans both devices on 142 qubits.
We discuss a path forward to create long-range gates and conclude in Sec.~\ref{sec:conclusion}.

\section{Circuit cutting\label{sec:cutting}}

Circuit cutting is a technique used to handle large quantum circuits that may not be directly executable on current quantum hardware due to limitations in qubit count or connectivity. 
This method involves decomposing a complex circuit into subcircuits that can be individually executed on quantum devices. 
The results from these subcircuits are then classically recombined to yield the result of the original circuit.

\begin{figure}[bp!]
    \centering
    \includegraphics[width=0.98\columnwidth]{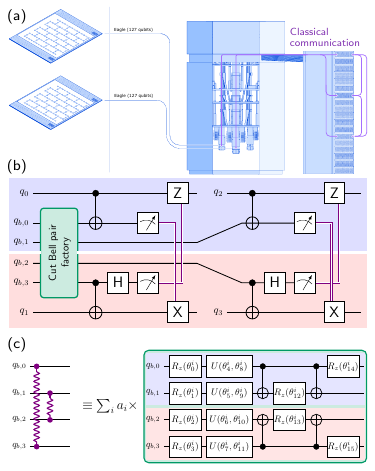}
    \caption{\textbf{Local operations and classical communication.} (a) Depiction of an IBM Quantum System Two architecture. Here, two 127 qubit Eagle QPUs are connected with a real-time classical link.
    (b) Quantum circuit to implement virtual CNOT gates on qubit pairs $(q_0, q_1)$ and $(q_2, q_3)$ with LOCC by consuming cut Bell pairs in a teleportation circuit.
    The purple double lines correspond to the real-time classical link.
    (c) Cut Bell pair factories $C_2(\boldsymbol{\theta}^i)$ for two simultaneously cut Bell pairs.
    The QPD has a total of 27 different parameter sets $\boldsymbol{\theta}^i$.
    Here, $U(\theta, \phi)=\sqrt{X} R_z(\theta)\sqrt{X}R_z(\phi)$.
    }
    \label{fig:locc}
\end{figure}

Mathematically, a single quantum channel $\mathcal{E}(\rho)$ is cut by expressing it as a sum over $I$ quantum channels $\mathcal{E}_i(\rho)$ resulting in the QPD
\begin{align}\label{eqn:qpd}
    \mathcal{E}(\rho)=\sum_{i=0}^{I-1} a_i\mathcal{E}_i(\rho).
\end{align}
The channels $\mathcal{E}_i(\rho)$ are easier to implement than $\mathcal{E}(\rho)$ and are built from LO~\cite{Mitarai2021} or LOCC~\cite{Piveteau2023}, see Fig.~\ref{fig:locc}.
Since some of the coefficients $a_i$ are negative, we introduce $\gamma=\sum_i|a_i|$ and $p_i= |a_i| / \gamma$ to recover a valid probability distribution over the channels $\mathcal{E}_i$ as
\begin{align}\label{eqn:qpd_gamma}
    \mathcal{E}(\rho)=\gamma\sum_i p_i {\rm sign}(a_i)\mathcal{E}_i(\rho).
\end{align}
Using this QPD we build an unbiased Monte-Carlo estimator $\hat{O}_\text{QPD}$ for $\mathrm{Tr}[O\mathcal{E}(\rho)]$ for an observable $O$ by replacing the $\mathcal{E}_i(\rho)$ in Eq.~(\ref{eqn:qpd_gamma}) with $\hat{O}_i$ which estimates $\mathrm{Tr}[O\mathcal{E}_i(\rho)]$.
Crucially, the cost of building $\hat{O}_\text{QPD}$ is a $\gamma^2$ increase in variance. 
To see this, assume that $\mathbb{E}[\hat{O}] = \gamma \mathbb{E}[\hat{O}_{\text{mix}}]$ and that the circuits all have a similar variance, i.e., $\mathrm{Var}(\hat{O})\approx \mathrm{Var}(\hat{O}_i)$. 
Here, $\hat{O}_{\text{mix}}$ can be seen as a probabilistic mixture of the random variables $\{\text{sign}(a_i)\hat{O}_i\}$, each chosen with probability $p_i$~\cite{Cai2023}.
Therefore, $\mathrm{Var}(\hat{O}_{\text{QPD}})=\gamma^2 \mathrm{Var}(\hat{O}_{\text{mix}})\approx \gamma^2\mathrm{Var}(\hat{O})$.
When cutting $n>1$ identical channels, we can build an estimator by taking the product of the QPDs for each individual channel, resulting in a $\gamma^{2n}$ rescaling factor~\cite{Temme2017,Endo2018}.
Such an increase in variance is compensated by a corresponding increase in the number of measured shots. 
Therefore, $\gamma^{2n}$ is called the \textit{sampling overhead}.
Circuit cutting is applicable to both wires~\cite{Peng2020,Brenner2023,Pednault2023} and gates~\cite{Hofmann2009, Mitarai2021, Piveteau2023} and the resulting quantum circuits have a similar structure.
We focus on gate cutting; details of the LO and LOCC quantum channels $\mathcal{E}_i$ and their coefficients $a_i$ are in Appendix~\ref{sec:lo} and \ref{sec:locc}, respectively.

Since implementing virtual gates with LOCC is a key contribution of our work we show how to create the required cut Bell pairs with local operations.
Here, multiple cut Bell pairs are engineered in a \emph{Cut Bell Pair Factory} since cutting multiple pairs at the same time requires a lower sampling overhead~\cite{Piveteau2023}, see Fig. \ref{fig:locc}(b) and \ref{fig:locc}(c).
Following Ref.~\cite{Vidal1999}, we first express the state $\rho_k$ of $k$ cut Bell pairs as the weighted sum of two separable states $\rho_k^\pm$, i.e., $\rho_k=(1+t)\rho_{k}^+-t\rho_k^-$ where $t$ is a non-negative finite real number.
Here, $\rho_k^\pm$ are separable over the disjoint qubit partitions $A$ and $B$ with cardinality $|A|=|B|=k$.
Next, to engineer $\rho_{k}^\pm$ we build a parametric quantum circuit $C_k(\boldsymbol{\theta}^i)$, i.e., the cut Bell pair factory, where no two-qubit gate connects qubits between $A$ and $B$. 
Both $\rho_{k}^\pm$ are mixed states that we realize via probabilistic mixtures of pure states $C_k(\boldsymbol{\theta}^i)$ with corresponding parameter sets $\boldsymbol{\theta}^i$ where $i$ corresponds to the QPD index of Eq.~(\ref{eqn:qpd}).
The sets of parameters $\boldsymbol{\theta}^i$ are classically optimized with SLSQP~\cite{Kraft1988} to minimize the $L_2$-norm between the statevectors of the target pure states and statevectors corresponding to the circuits $C_k(\boldsymbol{\theta}^i)$. The parametric circuit $C_k(\boldsymbol{\theta})$ is chosen such that the resulting error for each pure state is upper bounded by $10^{-8}$.
Crucially, to allow a rapid execution of the QPD with parametric updates all the parameters are the angles of virtual-$Z$ rotations~\cite{McKay2017}, see Fig.~\ref{fig:locc}(c).
Details are in Appendix~\ref{sec:bell_pair_qpds}.
Since the cut Bell pair factory forms two disjoint quantum circuits we place each sub-circuit close to qubits that have long-range gates.
The resulting resource is then consumed in a teleportation circuit.
For instance, in Fig.~\ref{fig:locc}(b) the cut Bell pairs are consumed to create CNOT gates on the qubit pairs (0, 1) and (2, 3).

\section{Periodic boundary conditions\label{sec:103nodes}}

\begin{figure*}
    \centering
    \includegraphics[clip, trim=5 5 5 5, width=0.98\textwidth]{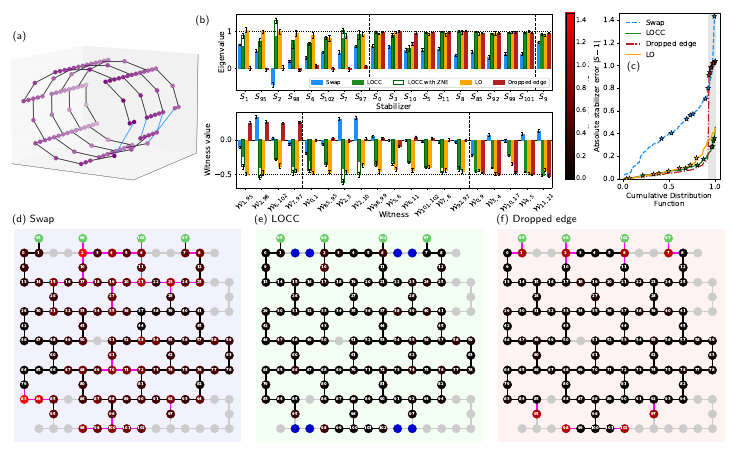}
    \caption{
    \textbf{Periodic boundary conditions.}
    (a) The heavy-hexagonal graph is folded on itself into a tubular form by the edges $(1, 95)$, $(2, 98)$, $(6, 102)$, and $(7, 97)$.
    (b) Top and bottom panels show the node stabilizers $S_j$ and witnesses $\mathcal{W}_{i,j}$, respectively, for the nodes and edges close to the long-range edges.
    (c) Cumulative distribution function of the stabilizer errors.
    The stars indicate node stabilizers $S_j$ that have an edge implemented by a long-range gate.
    In the dropped edge benchmark (dash-dotted red line) the long-range gates are not implemented and the star-indicated stabilizers thus have unit error.
    The grey region is the probability mass corresponding to node stabilizers affected by the cuts.
    In the two-dimensional layouts (d), (e) and (f) the green nodes duplicate nodes 95, 98, 102, and 97 to show the cut edges.
    The color of node $i$ is the absolute error $|S_i-1|$ of the measured stabilizer, as indicated by the colorbar.
    An edge is black if entanglement statistics are detected at a 99\% confidence level, and violet if not.
    In (d) the long range gates are implemented with SWAP gates.
    In (e) the same gates are implemented with LOCC.
    In (f) they are not implemented at all.
    }
    \label{fig:long_range}
\end{figure*}

As a first application of gate cutting, we construct a graph state with periodic boundary conditions on \emph{ibm\_kyiv}, an Eagle processor~\cite{Kim2023}, going beyond the limits imposed by its physical connectivity. 
A graph state $\ket{G}$, corresponding to a graph $G=(V,E)$ on nodes $V$ and edges $E$, is a quantum state on $|V|$ qubits created by initializing the qubits in the $\ket{+}$ state and subsequently applying controlled-$Z$ gates between qubits $(i,j)\in E$~\cite{Briegel2001,Hein2004}, see Appendix~\ref{sec:graph_states}.
Here, $G$ has $|V|=103$ nodes and requires four long-range edges $E_\text{lr}=\{(1, 95)$, $(2, 98)$, $(6, 102)$, $(7, 97)\}$ between the top and bottom qubits of the Eagle processor, see Fig.~\ref{fig:long_range}(a).
We measure the node stabilizers $S_i$ at each node $i\in V$ and the edge stabilizers formed by the product $S_iS_j$ across each edge $(i, j)\in E$.
From these stabilizers we build an entanglement witness $\mathcal{W}_{i,j}=(1-\langle S_i\rangle-\langle S_j\rangle-\langle S_iS_j\rangle)/4$ that is negative if there is bipartite entanglement across the edge $(i,j)\in E$~\cite{Zander2024}, see Appendix~\ref{sec:witness}.

We prepare $\ket{G}$ with three different methods.
The hardware native edges are always implemented with CNOT gates but
the periodic boundary conditions are implemented (i) with SWAP gates, (ii) with LOCC, and (iii) with LO to connect qubits across the whole lattice.
In (iii) we implement the CZ gate with six circuits built from $R_z$ gates and MCMs~\cite{Mitarai2021}.
Cutting four CZ gates with LO thus requires $I=6^4=1296$ circuits, see Appendix~\ref{sec:lo}.
However, since the node and edge stabilizers are at most in the light-cone~\cite{Tran2020} of one virtual gate we instead implement two QPDs in parallel which requires $I=6^2=36$ LO circuits per expectation value.
Turning to LOCC, we similarly construct two QPDs in parallel with $I=27$ circuits, each QPD implementing two long-range CZ gates.
The main difference with LO is a feed-forward operation consisting of single-qubit gates conditioned on $2n$ measurement outcomes, where $n$ is the number of cuts.
Each of the $2^{2n}$ cases triggers a unique combination of $X$ and/or $Z$ gates on the appropriate qubits.
Acquiring the measurement results, determining the corresponding case, and acting based on it is performed in real time by the control hardware, at the cost of a fixed added latency.
We mitigate and suppress the errors resulting from this latency with zero-noise extrapolation~\cite{Temme2017} and staggered dynamical decoupling~\cite{Viola1999, Mundada2023}, see Appendix~\ref{app:app_switch}.

\begin{table}[]
    \centering
    \begin{tabular}{l r r r r r} \hline\hline
         &  $\gamma^{2n}$ & Nbr. CNOTs & Nbr. MCM & $\sum_{i\in V} |S_i-1|$ \\ \hline
         Dropped edge & 1 & 112 & 0 & 13.1\\
         SWAPs & 1 & 374 & 0 & 44.3 \\
         LOCC & 49 & 128 & 8 & 8.9 \\
         LO & 81 & 112 & 8/3 & 7.0\\\hline\hline
    \end{tabular}
    \caption{
    \textbf{Circuit structure and node error.} 
    The circuits are transpiled to hardware-native CNOT gates. 
    The number of MCMs for LO varies with the different circuits in the QPD. 
    We therefore report the average number of MCMs.}
    \label{tab:execution_summary}
\end{table}

We benchmark the SWAP, LOCC, and LO implementations of $\ket{G}$ with a hardware-native graph state on $G'=(V,E')$ obtained by removing the long-range gates, i.e., $E'=E\backslash E_\text{lr}$.
The circuit preparing $\ket{G'}$ thus only requires 112 CNOT gates arranged in three layers following the heavy-hexagonal topology of the Eagle processor.
This circuit will report large errors when measuring the node and edge stabilizers of $\ket{G}$ for nodes on a cut gate since it is designed to implement $\ket{G'}$.
We refer to this hardware-native benchmark as the ``dropped edge benchmark''.
The swap-based circuit requires an additional 262 CNOT gates to create the long-range edges $E_\text{lr}$ which drastically reduces the value of the measured stabilizers, see Fig.~\ref{fig:long_range}(b), (c), and (d).
By contrast, the LOCC and LO implementation of the edges in $E_\text{lr}$ do not require SWAP gates.
The errors of their node and edge stabilizers for nodes not involved in a cut gate closely follow the dropped edge benchmark, see Fig.~\ref{fig:long_range}(b) and (c).
Conversely, the stabilizers involving a virtual gate have a lower error than the dropped edge benchmark and the swap implementation, see star markers in Fig.~\ref{fig:long_range}(c).
As overall quality metric we first report the sum of absolute errors on the node stabilizers, i.e., $\sum_{i\in V} |S_i-1|$, see Tab.~\ref{tab:execution_summary}.
The large SWAP overhead is responsible for the 44.3 sum absolute error.
The 13.1 error on the dropped edge benchmark is dominated by the eight nodes on the four cuts, see star markers Fig.~\ref{fig:long_range}(c).
By contrast, the LO and LOCC errors are affected by mid-circuit measurements.
We attribute the 1.9 additional error of LOCC over LO to the delays and the CNOT gates in the teleportation circuit and cut Bell pairs.
In the SWAP-based results $\mathcal{W}_{i,j}$ does not detect entanglement across 35 of the 116 edges at the 99\% confidence level, see Fig.~\ref{fig:long_range}(b) and (d).
For the LO and LOCC implementation $\mathcal{W}_{i,j}$ witnesses the statistics of bipartite entanglement across all edges in $G$ at the 99\% confidence level, see Fig.~\ref{fig:long_range}(e).
Finally, we observe that the variance of the stabilizers implemented with LO is on average $1.48(8)\times$ larger than those measured with LOCC, see Appendix~\ref{app:variance}.
This is in close agreement with the theoretically expected $1.65\times$ increase.
We attribute the difference to the increased noise in the dynamic circuits in LOCC.

\begin{figure*}[htbp!]
    \includegraphics[width=0.99\textwidth, clip, trim=0 3 0 0]{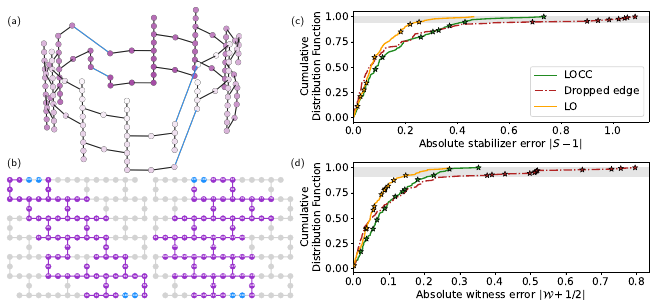}
    \caption{\textbf{Two connected QPUs with LOCC.} 
    (a) Graph state with periodic boundaries shown in 3D. 
    The blue edges are the cut edges.
    (b) Coupling map of two Eagle QPUs operated as a single device with 254 qubits. 
    The purple nodes are the qubits forming the graph state in (a) and the blue nodes are used for cut Bell pairs.
    Absolute error on the stabilizers (c) and edge witnesses (d) implemented with LOCC (solid green), LO (solid orange), and on a dropped-edge benchmark graph (dotted-dashed red) for the graph state in (a).
    In panels (c) and (d) the stars show stabilizers and edge witnesses that are affected by the cuts.
    In panels (c) and (d) the grey region is the probability mass corresponding to node stabilizers and edge witnesses, respectively, affected by the cut.
    In (c) and (d) we observe that the LO implementation outperforms the dropped-edge benchmark which we attribute to better device conditions as this data was taken on a different day from the benchmark and LOCC data.
    \label{fig:multi_qpu}}
\end{figure*}

\section{Connecting quantum processing units with LOCC\label{sec:134_nodes}}

We now combine two Eagle QPUs with 127 qubits each into a single QPU through a real-time classical connection.
Operating the devices as a single, larger processor consists of executing quantum circuits spanning the larger qubit register. 
In addition to unitary gates and measurements running concurrently on the merged QPU, we use dynamic circuits to perform gates that act on qubits on both devices. 
This is enabled by a tight synchronization and fast classical communication between physically separate instruments, required to collect measurement results and determine the control flow across the whole system~\cite{Gupta2023}. 

We test this real-time classical connection by engineering a graph state on 134 qubits built from heavy-hexagonal rings that snake through both QPUs, see Fig.~\ref{fig:multi_qpu}.
This graph forms a ring in three-dimensions, and requires four long-range gates which we implement with LO and LOCC.
As before, the LOCC protocol thus requires two additional qubits per cut gate for the cut Bell pairs.
Similarly to Sec.~\ref{sec:103nodes}, we benchmark our results to a graph that does not implement the edges that span both QPUs.
Crucially, since there is no quantum link between the two devices a benchmark with SWAP gates is impossible.
All edges exhibit the statistics of bipartite entanglement when we implement the graph with LO and LOCC at a 99\% confidence level.
Furthermore, the LO and LOCC stabilizers have the same quality as the dropped edge benchmark for nodes that are not affected by a long-range gate, see Fig.~\ref{fig:multi_qpu}(c).
Stabilizers affected by long-range gates have a large reduction in error compared to the dropped edge benchmark.
The sum of absolute errors on the node stabilizers $\sum_{i\in V} |S_i-1|$, is 21.0, 19.2, and 12.6 for the dropped edge benchmark, LOCC, and LO, respectively.
The LOCC results demonstrate how a dynamic quantum circuit in which two sub-circuits are connected by a real-time classical link can be executed on two otherwise disjoint QPUs.
The LO results could be obtained on a single device with 127 qubits at the cost of an additional factor of two in runtime since the sub-circuits can be run successively.

\section{Discussion and Conclusion\label{sec:conclusion}}

We implement long-range gates with LO and LOCC. With these gates, we engineer periodic boundary conditions on a 103-node planar lattice and connect in real time two Eagle processors to create a graph state on 134 qubits, going beyond the capabilities of a single chip.
Our cut Bell pair factories enable the LOCC scheme presented in Ref.~\cite{Piveteau2023}.
Both the LO and LOCC protocol deliver high-quality results that closely match a hardware native benchmark.
We observe that the larger $\gamma$ of the LO decomposition we use increases the variance of the stabilizers compared to the LOCC decomposition, see Appendix~\ref{app:variance}.
We attribute deviations from the theoretically expected $1.65\times$ increase to the long delay present in dynamic circuits during which error sources such as $T_1$, $T_2$ and static $ZZ$ cross-talk act.

The variance increase from the QPD is why research now focuses on reducing $\gamma$. 
It was recently shown that cutting multiple two-qubit gates \emph{in parallel} results in optimal LO QPDs with the same $\gamma$ as LOCC~\cite{Ufrecht2023, Schmitt2023}.
Crucially, in LOCC the QPD is only required to cut the Bell pairs.
This costly QPD could be removed, i.e., no shot overhead, by distributing entanglement across multiple chips~\cite{Cirac1997, Duan2001}.
In the near to medium term this could be done by operating gates in the microwave regime over conventional cables~\cite{Magnard2020, Zhong2021, Niu2023} or, alternatively in the long term, with an optical-to-microwave transduction~\cite{Orcutt2020, Lauk2020, Krastanov2021}.
Entanglement distribution is typically noisy and may result in non-maximally entangled states.
However, gate teleportation requires a maximally entangled resource.
Nevertheless, non-maximally entangled states could lower the sampling cost of the QPD~\cite{Bechtold2023} and multiple copies of non-maximally entangled states could be distilled into a pure state for teleportation~\cite{Bennett1996} either during the execution of a quantum circuit or possibly during the delays between consecutive shots, which may be as large as $250~\mathrm{\mu s}$ for resets~\cite{Wack2021, Tornow2022}. 
Combined with these settings, our error-mitigated and suppressed dynamic circuits would enable a modular quantum computing architecture without the sampling overhead of circuit cutting.

Our real-time classical link implements long-range gates and classically couples disjoint quantum processors.
Crucially, the cut Bell pairs that we present have value beyond our work.
For example, they are directly usable to cut circuits in measurement-based quantum computing~\cite{Baumer2023}.
Furthermore, the combination of staggered dynamical decoupling with zero-noise extrapolation mitigate the lengthy delays of the feed-forward operations which enables a high-quality implementation of dynamic circuits.
Our work sheds light into the noise sources that a transpiler for distributed superconducting quantum computers must consider~\cite{Caleffi2022}.
In summary, error-mitigated dynamic circuits with a real-time classical link across multiple QPUs enable a modular scaling of quantum computers.

\section{Acknowledgements}

We acknowledge the use of IBM Quantum services for this work. 
The views expressed are those of the authors, and do not reflect the official policy or position of IBM or the IBM Quantum team.
We acknowledge the work of the IBM Quantum software and hardware teams that enabled this project.
In particular, Brian Donovan, Ian Hincks, and Kit Barton for circuit compilation and execution. 
We also thank David Sutter, Jason Orcutt, Ewout van den Berg, and Paul Seidler for useful discussions. 

\appendix

\begin{figure*}
    \includegraphics[width=\textwidth]{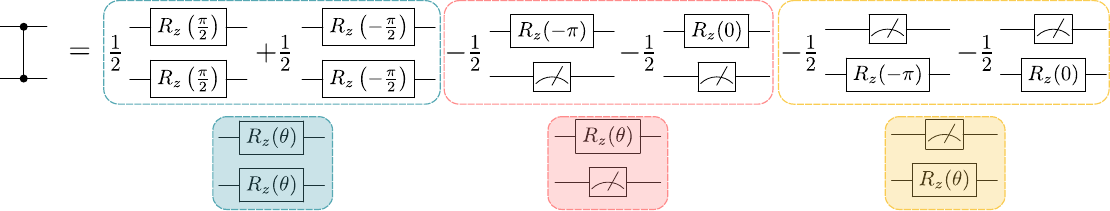}
    \caption{\textbf{LO decomposition of a $\mathrm{CZ}$ gate.} 
    A single $\mathrm{CZ}$ gate can be simulated through local operations by sampling from the shown QPD and applying classical post-processing to the results. 
    Each of the six circuits has a sampling probability of $1/(2\gamma)=1/6$. 
    For the four circuits featuring mid-circuit measurements, the corresponding QPD coefficient is adjusted by a factor of $+1$ for outcome 0 and a factor of $-1$ for outcome 1. 
    To optimize the execution, these six circuits are consolidated into three parametrized circuits to enable a parametric circuit execution. 
    \label{fig:cz_lo}}
\end{figure*}

\section{Virtual gates implemented with LO\label{sec:lo}}

This appendix discusses how we implement virtual $\mathrm{CZ}$ gates with LO by following Ref.~\cite{Mitarai2021}. 
We therefore decompose each cut $\mathrm{CZ}$ gate into local operations and a sum over six different circuits defined by
\begin{align}
\begin{split}
    2\mathrm{CZ} &= \sum_{\alpha\in\{\pm 1\}}R_z\left(\alpha\frac{\pi}{2}\right)\otimes R_z\left(\alpha\frac{\pi}{2}\right)\\
    & - \sum_{\alpha_1, \alpha_2\in\{\pm 1\}} \alpha_1 \alpha_2 R_z\left(-\frac{\alpha_1 + 1}{2}\pi\right) \otimes \left(\frac{I + \alpha_2 Z}{2}\right)\\
    & - \sum_{\alpha_1, \alpha_2\in\{\pm 1\}}\alpha_1 \alpha_2 \left(\frac{I + \alpha_1 Z}{2}\right) \otimes R_z\left(-\frac{\alpha_2 + 1}{2}\pi\right).
\end{split}\label{eqn:cz_lo_decomposition}
\end{align}
Here, $R_z(\theta)=\exp\left(-i\frac{\theta}{2}Z\right)$ are virtual Z rotations~\cite{McKay2017}.
The factor 2 in front of $\mathrm{CZ}$ is for readability. 
Each of the possible six circuits is thus weigthed by a $1/6$ probability, see Fig.~\ref{fig:cz_lo}.
The operations $\left(I + Z\right)/2$ and $\left(I - Z\right)/2$ correspond to the projectors $\ket{0}\!\!\bra{0}$ and $\ket{1}\!\!\bra{1}$, respectively.
They are implemented by mid-circuit measurements and classical post-processing. 
More specifically, when computing the expectation value of an observable $\langle O\rangle=\sum_i a_i\langle O\rangle_i$ with the LO QPD, we multiply the expectation values $\langle O\rangle_i$ by $1$ and $-1$ when the outcome of a mid-circuit measurement is $0$ and $1$, respectively.

In general, sampling from a QPD results in an overhead of $(\sum_{i=0}^{I-1}|a_i|)^2$ where $I$ is the number of circuits in the QPD and the $a_i$ are the QPD coefficients~\cite{Cai2023}.
However, since the LO QPDs in our experiments only have 36 circuits we fully enumerate the QPDs by executing all 36 circuits.
The sampling cost of full enumeration is $I(\sum_{i=0}^{I-1}|a_i|^2)$.
Furthermore, since $|a_i|=1/2~\forall~i=0,...,I-1$ sampling from the QPD and fully enumerating it both have the same shot overhead.

The decomposition in Eq.~(\ref{eqn:cz_lo_decomposition}) with $\gamma^2=9$ is optimal with respect to the sampling overhead for a \textit{single} gate~\cite{Piveteau2023}.
Recently, Refs.~\cite{Ufrecht2023, Schmitt2023} found a new protocol that achieves the same $\gamma$ overhead as LOCC when cutting multiple gates in parallel.
The proofs in Refs.~\cite{Ufrecht2023, Schmitt2023} are theoretical demonstrating the existence of a decomposition.
However, finding a circuit construction to implement the underlying QPD is still an outstanding task.

\section{Virtual gates implemented with LOCC\label{sec:locc}}

We now discuss the implementation of the dynamic circuits that enable the virtual gates with LOCC.
We first present an error suppression and mitigation of dynamic circuits with dynamical decoupling (DD) and zero-noise extrapolation (ZNE).
Second, we discuss the methodology to create the cut Bell pairs and present the circuits to implement one, two, and three cut Bell pairs.
Finally, we propose a simple benchmark experiment to assess the quality of a virtual gate.

\subsection{Error mitigated quantum circuit switch instructions\label{app:app_switch}}

\begin{figure}[htbp]
    \centering
    \includegraphics[width=\columnwidth]{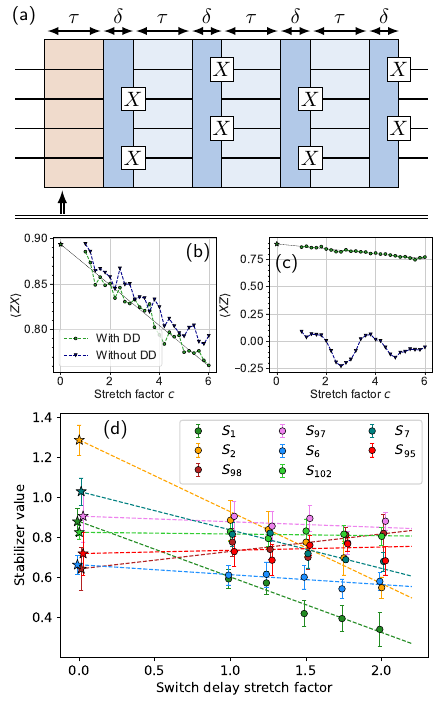}
    \caption{
    \textbf{Zero-noise extrapolation of a switch.}
    (a)~Implementation of a switch with DD.
    The conditional gates (not shown) are executed after the last DD $X$ gate.
    The red delay of $\tau$ shows the duration in which no gates can be executed as the control electronics is busy, see Sec.~\ref{app:app_switch}.
    The three additional delays of $\tau$ enable staggered DD.
    The four additional and variable delays of $\delta$ allow us to vary the duration of the switch for ZNE.
    Panels (b) and (c) show the $ZX$ and $XZ$ correlators measured on \emph{ibm\_peekskill} as a function of the switch stretch factor $c$ for a two-qubit graph state on $G=(\{0, 1\}, \{(0, 1)\})$.
    (d) Example correlators of the 103 node graph extrapolated with ZNE.
    }
    \label{fig:switch_zne}
\end{figure}

All quantum circuits presented in this work are written in Qiskit.
The feed-forward operations of the LOCC circuits are executed with a quantum circuit switch instruction, hereafter referred to as a \emph{switch}. 
A switch defines a set of cases where the quantum circuit can branch to, depending on the outcome of a corresponding set of measurements.
This branching occurs in real time for each experimental shot, with the measurement outcomes being collected by a central processor, which in turn broadcasts the selected case (here corresponding to a combination of $X$ and $Z$ gates) to all control instruments. 
This process results in a latency of the order of $0.5~\mu\mathrm{s}$ (independent of the selected case) during which no gates can be applied, red area in Fig.~\ref{fig:switch_zne}(a).
Free evolution during this period ($\tau$), often dominated by static $ZZ$ crosstalk in the Hamiltonian, typically with a strength ranging from $\sim 10^3$ to $10^4 ~ \mathrm{Hz}$, significantly deteriorates results. 
To cancel this unwanted interaction and any other constant or slowly fluctuating $IZ$ or $ZI$ terms, we precede the conditional gates with a staggered DD $X$-$X$ sequence, adding $3\tau$ to the switch duration, see Fig.~\ref{fig:switch_zne}(a). 
The value of $\tau$ is determined by the longest latency path from one QPU to the other and is fine tuned by maximizing the signal on such a DD sequence.
Furthermore, we mitigate the effect of the overall delay on the observables of interest with ZNE~\cite{Temme2017}.
To do this, we first stretch the switch duration by a factor $c=(\tau+\delta)/\tau$, where $\delta$ is a variable delay added before each $X$ gate in the DD sequence, see Fig.~\ref{fig:switch_zne}(a).
Second, we extrapolate the stabilizer values to the zero-delay limit $c=0$ with a linear fit.
In many cases an exponential fit can be justified~\cite{Kim2023}, however, we observe in our benchmark experiments that a linear fit is appropriate, see Fig.~\ref{fig:switch_zne}.
Crucially, without DD we observe strong oscillations in the measured stabilizers which prevent an accurate ZNE, see the $XZ$ stabilizer in Fig.~\ref{fig:switch_zne}(c).
As seen in the main text, this error suppression and mitigation reduces the error on the stabilizers affected by virtual gates.

The error suppression and mitigation that we implement for the switch also applies to other control flow statements.
Indeed, the switch is not the only instruction capable of representing control flow.
For instance, OpenQASM3~\cite{Cross2022} supports if/else statements.
Our scheme is done by (i) adding DD sequences to the latency (possibly by adding delays if the control electronics cannot emit pulses during the latency), (ii) stretching the delay, and (iii) extrapolating to the zero-delay limit.

\subsection{Cut Bell pair factories\label{sec:bell_pair_qpds}}

Here, we discuss the quantum circuits to prepare the cut Bell pairs needed to realize virtual gates with LOCC.
To create a factory for $k$ cut Bell pairs we must find a linear combination of circuits with two disjoint partitions with $k$ qubits each to reproduce the statistics of Bell pairs.
We create the state $\rho_k$ of the Bell pairs following Ref.~\cite{Vidal1999} such that $\rho_k=(1+t_k)\rho_{k}^+-t_k\rho_k^-$ where $t_k=2^k-1$.
Here, $\rho_k^\pm$ are mixed states separable on each partition. 
The total cost of this QPD with two states is determined by $\gamma_k=2t_k+1$.
Next, we realize $\rho_k^\pm$ from a probabilistic mixture of pure states $\rho_{k,i}^\pm$, i.e., valid probability distributions.
Following Ref.~\cite{Vidal1999}, we need $\smash{n_k^+=2^{2^k}-1}$ pure states to realize $\rho_k^+$.
The exact form of $\rho_k^+$, omitted here for brevity, is given in Appendix B of Ref.~\cite{Vidal1999}.
We implement $\rho_k^-$, a diagonal density matrix of all basis states that do not appear in $\rho_k$,  with $n_k^-=4^k-2^k$ basis states.
Therefore, the total number of parameter sets $I=n_k^++n_k^-$ required to implement one, two, and three cut Bell pairs is 5, 27, and 311, respectively.
Finally, the coefficients $a_{i,k}$ of all the circuits in the QPD in Eq.~(\ref{eqn:qpd}) that implement $\rho_k^\pm$ are
\begin{align}
    a_{i,k}=&\frac{1+t_k}{n_k^+},~\text{for}~i\in\{0,...,n_k^+-1\}\text{, and} \\
    a_{i,k}=&-\frac{t_k}{n_k^-},~\text{for}~i\in\{n_k^+,...,n_k^++n_k^--1\}.
\end{align}
For $k=2$ the resulting weights, $|a_{i,k}|/\gamma_k$ are approximately all equal. 
There is thus no practical difference between sampling and enumerating the $k=2$ QPD when executing it on hardware.
More precisely, for the factories with two cut Bell pairs that we run on hardware, the cost of sampling the QPD is $(\sum_{i=0}^{I-1}|a_{i,2}|)^2=\gamma_2^2(1+1.6\cdot 10^{-7})$ and the cost of fully enumerating the QPD is $I(\sum_{i=0}^{I-1}|a_{i,2}|^2)=\gamma_2^2(1+1.0\cdot 10^{-3})$ where $\gamma_2=7$.

\begin{figure}[htbp!]
    \centering
    \includegraphics[width=\columnwidth]{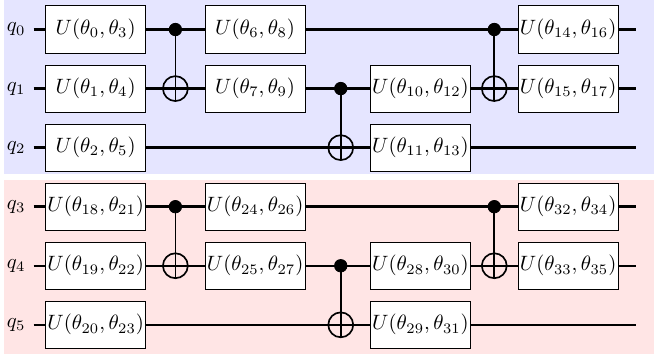}
    \caption{\textbf{Quantum circuit of three cut Bell pairs.}
    A sum over the right set of parameter vectors $\smash{\{\boldsymbol{\theta}\}}$ results in three cut Bell pairs between qubit pairs $(q_0, q_3)$, $(q_1, q_4)$, and $(q_2, q_5)$.
    The gate $U(\theta,\phi)$ corresponds to the gate sequence $\sqrt{X}R_z(\theta)\sqrt{X}R_z(\phi)$.
    The blue and red shaded regions correspond to the two disjoint portions of the quantum circuit.}
    \label{fig:triple_bp}
\end{figure}

We construct all pure states $\rho_{k,i}^\pm$ from the same template variational quantum circuit $\smash{C_k(\boldsymbol{\theta}^i)}$ with parameters $\boldsymbol{\theta}^i$ where the index $i=0, ..., I-1$ runs over the $I$ elements of the probabilistic mixtures defining $\rho_k^\pm$.
The parameters $\smash{\boldsymbol{\theta}^i}$ in the template circuits $\smash{C_k(\boldsymbol{\theta}^i)}$ are optimized by the SLSQP classical optimizer by minimizing the $L_2$-norm with respect to the $I$ pure target states needed to represent $\rho_k^\pm$, where the norm is evaluated with a classical statevector simulation. 
We test different Ans\"atze and find that the ones provided in Figs.~\ref{fig:locc}(c) and \ref{fig:triple_bp} allow to achieve an error, based on the $L_2$-norm, of less than $10^{-8}$ for each state while having minimal hardware requirements.
Note that for $\rho_k^-$, we can analytically derive the parameters and could significantly simplify the Ansatz. 
However, we keep the same template for compilation and execution efficiency.

A single cut Bell pair is engineered by applying the gates $U(\theta_0, \theta_1)$ and $U(\theta_2, \theta_3)$ on qubits~0 and~1.
Here, and in the figures, the gate $U(\theta,\phi)$ corresponds to $\sqrt{X}R_z(\theta)\sqrt{X}R_z(\phi)$.
The QPD of a single cut Bell pair requires five sets of parameters given by $\{[\pi/2, 0, \pi/2, 0],$ $[\pi/2, -2\pi/3, \pi/2, 2\pi/3],$ $[\pi/2, 2\pi/3,$ $\pi/2, -2\pi/3],$ $[\pi, 0, 0, 0],$ $[0, 0, \pi, 0]\}$.
The circuits to simultaneously create two and three cut Bell pairs are shown in Fig.~\ref{fig:locc}(c) of the main text and Fig.~\ref{fig:triple_bp}, respectively.
The values of these parameters and the circuits can be made available on reasonable request.

\subsection{Benchmarking qubits for LOCC}

\begin{figure}
    \centering
    \includegraphics[width=\columnwidth]{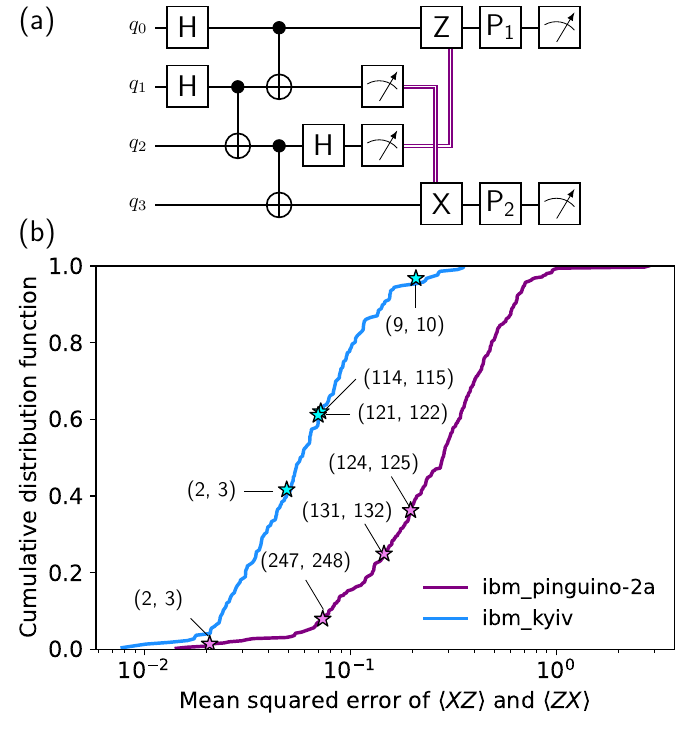}
    \caption{
    \textbf{LOCC Bell pair benchmark.}
    (a) Quantum circuit that creates an uncut Bell pair on qubits (1, 2) and consume it in a teleportation circuit to create a Bell state on qubits (0, 3).
    (b) Cumulative distribution function of the MSE of $\langle ZX\rangle$ and $\langle XZ\rangle$ for all groups of four linearly connected qubits on each device.
    The stars correspond to the qubits used in the 103- and 134-node graph states presented in the main text.
    The numbers in brackets indicate the qubit numbers corresponding to $(q_1, q_2)$ in panel~(a).
    }
    \label{fig:bell_benchmark}
\end{figure}

The quality of a CNOT gate implemented with dynamic circuits depends on hardware properties.
For example, qubit relaxation, dephasing, and static $ZZ$ crosstalk all negatively impact the qubits during the idle time of the switch.
Furthermore, measurement quality also impacts virtual gates implemented with LOCC.
Indeed, errors on mid-circuit measurements are harder to correct than errors on final measurements as they propagate to the rest of the circuit through the conditional gates~\cite{Gupta2023b}. 
For instance, assignment errors during readout results in an incorrect application of a single-qubit $X$ or $Z$ gate. 
Given the variability in these qubit properties, care must be taken in selecting those to act as cut Bell pairs. 
To determine which qubits will perform well as cut Bell pairs we develop a fast characterization experiment on four qubits that does not require a QPD or error mitigation. 
This experiment creates a graph state between qubits 0 and 3 by consuming an uncut Bell pair created on qubits 1 and 2 with a Hadamard and a CNOT gate. 
We measure the stabilizers $ZX$ and $XZ$ which require two different measurement bases. 
The resulting circuit, shown in Fig.~\ref{fig:bell_benchmark}(a), is structurally equivalent to half of the circuit that consumes two cut Bell pairs, e.g., Fig.~\ref{fig:locc}(c) of the main text.
We execute this experiment on all qubit chains of length four on the devices that we use and report the mean squared error (MSE), i.e., $[(\langle ZX\rangle-1)^2 + (\langle XZ\rangle-1)^2]/2$ as a quality metric. 
The lower the MSE is the better the set of qubits act as cut Bell pairs.
With this experiment we benchmark, \emph{ibm\_kyiv} (the device used to create the graph state with 103 nodes), and \emph{ibm\_pinguino-1a} and \emph{ibm\_pinguino-1b} (the two Eagle QPUs combined into a single device, named \emph{ibm\_pinguino-2a}, used to create the graph state with 134 nodes). 
We observe more than an order of magnitude variation in MSE across each device, see Fig.~\ref{fig:bell_benchmark}(b).

Crucially, the qubits we chose to act as cut Bell pairs is a tradeoff between the graph we want to engineer and the quality of the MSE benchmark. 
For example, the graphs with periodic boundary conditions presented in the main text were designed first based on the desired shape of $\ket{G}$ and second based on the MSE of the Bell pair quality test.

\section{Graph states\label{sec:graph_states}}

A graph state $\ket{G}$ is created from a graph $G=(V, E)$ with nodes $V$ and edges $E$ by applying an initial Hadamard gate to each qubit, corresponding to a node in $V$, and then $\mathrm{CZ}$ gates to each pair of qubits $(i, j)\in E$.
The resulting state $\ket{G}$ has $|V|$ first-order stabilizers, one for each node $i\in V$, defined by $S_i=X_i\prod_{k\in N(i)}Z_k$. 
Here, $N(i)$ is the neighborhood of node $i$ defined by $E$.
These stabilizers satisfy $S_i\ket{G}=\ket{G}$.
By construction, any product of stabilizers is also a stabilizer.
If an edge $(i,j)\in E$ is not implemented by a $\rm CZ$ gate the corresponding stabilizers drop to zero, i.e., $\expval{S_i}=\expval{S_j}=0$.
This effect can be seen in the dropped edge benchmark, see, e.g., Fig.~\ref{fig:long_range}(b) of the main text.

\section{Entanglement Witness\label{sec:witness}}

We now define a success criterion for the implementation of a graph state with entanglement witnesses~\cite{Jungnitsch2011}.
A witness $\mathcal{W}$ is designed to detect a certain form of entanglement.
Since we cut edges in the graph state we focus on witnesses $\mathcal{W}_{i,j}$ over two nodes $i$ and $j$ connected by an edge in $E$.
An edge $(i, j)$ of our graph state $\ket{G}$ presents entanglement if the expectation-value $\langle\mathcal{W}_{i,j}\rangle<0$.
The witness does not detect entanglement if $\langle\mathcal{W}_{i,j}\rangle\geq0$.
The first-order stabilizers of nodes $i$ and $j$ with $(i,j)\in E$ are
\begin{align}
  S_i=Z_jX_i\!\!\!\!\prod_{k\in N(i)\setminus j}\!\!\!\!Z_k\quad\text{and}\quad S_j=X_jZ_i\!\!\!\!\prod_{k\in N(j)\setminus i}\!\!\!\!Z_k  .
\end{align}
Here, $N(i)$ is the neighborhood of node $i$ which includes $j$ since $(i,j)\in E$.
Thus, $N(i)\setminus j$ is the neighborhood of node $i$ excluding $j$.
Following Refs.~\cite{Toth2005, Jungnitsch2011}, we build an entanglement witness for edge $(i,j)\in E$ as
\begin{align}
  \mathcal{W}_{i,j}=\frac{1}{4}\mathbb{I}-\frac{1}{4}\left(\expval{S_i}+\expval{S_j}+\expval{S_iS_j}\right).
\end{align}
This witness is zero or positive if the states are separable.
Alternatively, as done in Ref.~\cite{Zander2024}, a witness for bi-separability is also given by
\begin{align}
    \mathcal{W}'_{i,j}=\mathbb{I}-\expval{S_i}-\expval{S_j}.
\end{align}
Here, we consider both witnesses. 
The data in the main text are presented for $\mathcal{W}_{i,j}$.
As discussed in Ref.~\cite{Toth2005}, $\mathcal{W}_{i,j}$ is more robust to noise than $\mathcal{W}'_{i,j}$. 
However, $\mathcal{W}_{i,j}$ requires more experimental effort to measure than $\mathcal{W}'_{i,j}$ due to the stabilizer $S_iS_j$.

For completeness we now show how a witness can detect entanglement by focusing on $\mathcal{W}_{i,j}$.
A separable state satisfies $\expval{P_1...P_n}=\prod_i\expval{P_i}$ where $P_i$ are single-qubit Pauli operators.
Therefore we can show, using the Cauchy-Schwarz inequality, that $\expval{S_i}+\expval{S_j}+\expval{S_iS_j}\leq1$ and that $\mathcal{W}_{i,j}\geq0$ for separable states.
\begin{align}
    \expval{S_i}+\expval{S_j}+\expval{S_iS_j} = \expval{Z_j}\expval{X_i}\!\!\!\!\prod_{k\in N(i)\setminus j}\!\!\!\!\expval{Z_k} \\ \label{eqn:w1}
    + \expval{X_j}\expval{Z_i}\!\!\!\!\prod_{k\in N(j)\setminus i}\!\!\!\!\expval{Z_k}+\expval{Y_i}\expval{Y_j}\!\!\!\!\prod_{k\in M(i,j)}\!\!\!\!\expval{Z_k} \\ \label{eqn:w2}
    \leq \left|\expval{Z_j}\right|\left|\expval{X_i}\right|+\left|\expval{X_j}\right|\left|\expval{Z_i}\right|+\left|\expval{Y_j}\right|\left|\expval{Y_i}\right| \\ \label{eqn:w3}
    \leq \sqrt{\expval{X_i}^2+\expval{Y_i}^2+\expval{Z_i}^2}\sqrt{\expval{X_j}^2+\expval{Y_j}^2+\expval{Z_j}^2} \\
    \leq 1
\end{align}
The step from Eq.~(\ref{eqn:w1}) to Eq.~(\ref{eqn:w2}) relies on $\prod_i a_i\leq \prod_i |a_i|$ and that $\prod_{k}|\expval{Z_k}|\leq 1$ where the product runs over nodes that do not contain $i$ or $j$.
The step from Eq.~(\ref{eqn:w2}) to Eq.~(\ref{eqn:w3}) is based on the Cauchy-Schwarz inequality.
The final step relies on the fact that $\expval{X_i}^2+\expval{Y_i}^2+\expval{Z_i}^2\leq 1$ with pure states being equal to one.
Therefore, the witness $\mathcal{W}_{i,j}$ will be negative if the state is not separable.

In the graph states presented in the main text we execute a statistical test at a 99\% confidence level to detect entanglement.
As discussed in Appendix~\ref{app:stab_analysis}, and shown in Fig.~\ref{fig:long_range}(b) of the main text, some witnesses may go below $-1/2$ due to readout error mitigation, the QPD, and Switch ZNE.
We therefore consider an edge to have the statistics of entanglement if the deviation from $-1/2$ is not statistically greater than $\pm1/2$.
Based on a one-tailed test, we consider that edge $(i, j)$ is bi-partite entangled if
\begin{align}\label{eqn:criteria1}
    -\frac{1}{2}+\left|\langle\mathcal{W}_{i,j}\rangle+\frac{1}{2}\right|+z_{99\%}\sigma_{\mathcal{W},i,j}<0.
\end{align}
Similarly, we form a success criterion based on $\mathcal{W}'_{i,j}$ as
\begin{align}\label{eqn:criteria2}
    -1+\left|\langle\mathcal{W}'_{i,j}\rangle+1\right|+z_{99\%}\sigma_{\mathcal{W}',i,j}<0.
\end{align}
This criterion penalizes any deviation from -1, i.e., the most negative value that $\mathcal{W}'_{i,j}$ can have.
Here, $z_{99\%}=2.326$ is the $z$-score of a Gaussian distribution at a 99\% confidence level and $\sigma_{\mathcal{W},i,j}$ is the standard deviation of edge witness $\mathcal{W}_{i,j}$.
These tests are conservative as they penalize any deviation from the ideal values.
In addition, these tests are most suitable for circuit cutting since the QPD may increase the variance $\sigma_{\mathcal{W}_{i,j}}$ of the measured witnesses.
Therefore, the statistics of entanglement are only detected if the mean of a witness is sufficiently negative and its standard deviation is sufficiently small.
An edge $(i,j)\in E$ fails the criteria if Eq.~(\ref{eqn:criteria1}) or Eq.~(\ref{eqn:criteria2}) is not satisfied.
All edges in $E$, including the cut edges, pass the test based on $\mathcal{W}_{i,j}$ when implemented with LO and LOCC, see Tab.~\ref{tab:pass_rates}.
However, some edges fail the test based on $\mathcal{W}'_{i,j}$ due to the lower noise robustness of $\mathcal{W}'_{i,j}$ compared to $\mathcal{W}_{i,j}$.

\begin{table}[]
    \centering
    \begin{tabular}{l r r r r} \hline\hline
         Graph & \multicolumn{2}{c}{103 nodes} & \multicolumn{2}{c}{134 nodes} \\
         Criteria & $\mathcal{W}_{i,j}$ (\ref{eqn:criteria1}) & $\mathcal{W}'_{i,j}$ (\ref{eqn:criteria2}) & $\mathcal{W}_{i,j}$ (\ref{eqn:criteria1}) & $\mathcal{W}'_{i,j}$ (\ref{eqn:criteria2})\\ \hline
         SWAP & 70\% & 48\% & n.a. & n.a. \\
         Dropped-edge & 89\% & 88\% & 92\% & 85\% \\
         LOCC & 100\% & 100\% & 100\% & 87\% \\
         LO   & 100\% & 100\% & 100\% & 96\% \\ \hline\hline
    \end{tabular}
    \caption{
    \textbf{Witness tests.} Fraction of the edges in the graph state that pass the entanglement witness tests.
    For the dropped-edge benchmark we expect a pass rate of at most 88\% and 92\% for the 103- and 134-node graphs respectively.
    The 89\% measured pass rate of dropped-edge for the graph state with 103 nodes exceeds this value due to a single edge that barely passes the test with $\mathcal{W}_{i,j}=-0.0917$ and $\sigma_{\mathcal{W},i,j}=0.0146$ due to measurement fluctuations.
    }
    \label{tab:pass_rates}
\end{table}

\section{Circuit count for stabilizer measurements}

Obtaining the bipartite entanglement witnesses requires measuring the expectation values of $\langle S_i\rangle$, $\langle S_j\rangle$ and $\langle S_iS_j\rangle$ of each edge $(i,j)\in E$.
For the 103- and 134-node graphs presented in the main text all 219 and 278 node and edge stabilizers, respectively, can be measured in $N_{\!_S}=7$ groups of commuting observables.
To mitigate final measurement readout errors we employ twirled readout error extinction (TREX) with $N_{\!_{\text{TREX}}}$ samples~\cite{Berg2022}.
When virtual gates are employed with LO and LOCC we require $I_{\rm LO}$ and \smash{$I_{\rm LOCC}$}
more circuits, respectively.
In this work, we fully enumerate the QPD.
Furthermore, for LOCC we mitigate the delay of the switch instruction with ZNE based on $N_{\!_{\text{ZNE}}}$ stretch factors.
Therefore, the four types of experiments are executed with the following number of circuits.
\begin{center}
\begin{tabular}{l l l}
    Swaps & & $N_SN_{\!_{\text{TREX}}}$ \\[0.3em]
    Dropped edge & & $N_SN_{\!_{\text{TREX}}}$ \\[0.3em]
    LO & & $N_SN_{\!_{\text{TREX}}}I_{\rm LO}$ \\[0.3em]
    LOCC & & $N_SN_{\!_{\text{TREX}}}I_{\rm LOCC}N_{\!_{\text{ZNE}}}$\\
\end{tabular}
\end{center}
In all experiments, we use $N_{\!_{\text{TREX}}}=5$ TREX samples.
Therefore, measuring the stabilizers without a QPD requires $N_\text{S}\times N_\text{TREX}=35$ circuits.
For LO and LOCC measuring the stabilizers for the graphs in the main text requires $6^4$ and $27^2$ circuits, respectively.
However, due to the graph structure each edge witness is only ever in the light-cone of two cut gates at most.
We may thus execute a total of $I_{\rm LO}=6^2$ and $I_{\rm LOCC}=27$ circuits for LO and LOCC, respectively, based on the light-cone of the gates.
For higher-weight observables this corresponds to sampling the diagonal terms of a joint QPD.
Therefore, measuring the stabilizers with LO requires $N_\text{S}\times N_\text{TREX}\times I_{\text{LO}}=1260$ circuits.
For LOCC we further error mitigate the switch with $N_{\!_{\text{ZNE}}}=5$ stretch factors.
We therefore execute $N_\text{S}\times N_\text{TREX}\times I_{\text{LOCC}}\times N_\text{ZNE}=4725$ circuits to measure the error mitigated stabilizers needed to compute $\mathcal{W}_{i,j}$.
Each circuit is executed with a total of 1024 shots.

To reconstruct the value of the measured observables, we first merge the shots from the TREX samples.
To do this we flip the classical bits in the measured bit-strings corresponding to measurements where TREX prepended an $X$ gate.
These processed bit-strings are then aggregated in a count dictionary with $1024\times N_\text{TREX}$ counts.
Next, to obtain the value of a stabilizer we identify which of the $N_S$ measurement basis we need to use.
The value of a stabilizer and its corresponding standard deviation are then obtained by resampling the corresponding $1024\times N_\text{TREX}$ counts.
Here, we randomly select 10\% of the shots to compute an expectation value.
Ten such expectation values are averaged and reported as the measured stabilizer value.
The standard deviation of these ten measurements is reported as the standard deviation of the stabilizer, shown as error bars in Fig.~\ref{fig:long_range}(b).
Finally, if the stabilizer is in the light-cone of a virtual gate implemented with LOCC we linearly fit the value of the stabilizer obtained at the $N_{\!_{\text{ZNE}}}=5$ switch stretch factors.
This fit, exemplified in Fig.~\ref{fig:switch_zne}(d), allows us to report the stabilizer at the extrapolated zero-delay switch.

\section{Stabilizer analysis and error mitigation\label{app:stab_analysis}}

The expectation value of a stabilizer $S$ computed from a set of bitstrings should always be contained in $[-1, 1]$.
A careful inspection of the stabilizers reported in the main text reveals values exceeding one.
This is due to the error mitigation and the QPD.
We first consider the ``dropped edge'' data in Fig.~\ref{fig:multi_qpu} in the main text which is not built from a QPD.
The TREX error mitigated stabilizers are computed by (i) twirling the readout with $X$ gates, (ii) merging the counts by flipping the measured bits for which TREX added an $X$ gate before the measurements, and (iii) computing $\langle S\rangle_\text{Twirl}/\langle S'\rangle_\text{TREX}$.
Here, $\langle S\rangle_\text{Twirl}$ is computed from the counts obtained in step (ii).
The denominator $\langle S'\rangle_\text{TREX}$ is the expectation value of $S'$ on the TREX calibration counts where $S'$ has a $Z$ operator for each non-identity Pauli in $S$.

\begin{figure}[htbp!]
    \centering
    \includegraphics[width=\columnwidth, clip, trim=6 0 5 0]{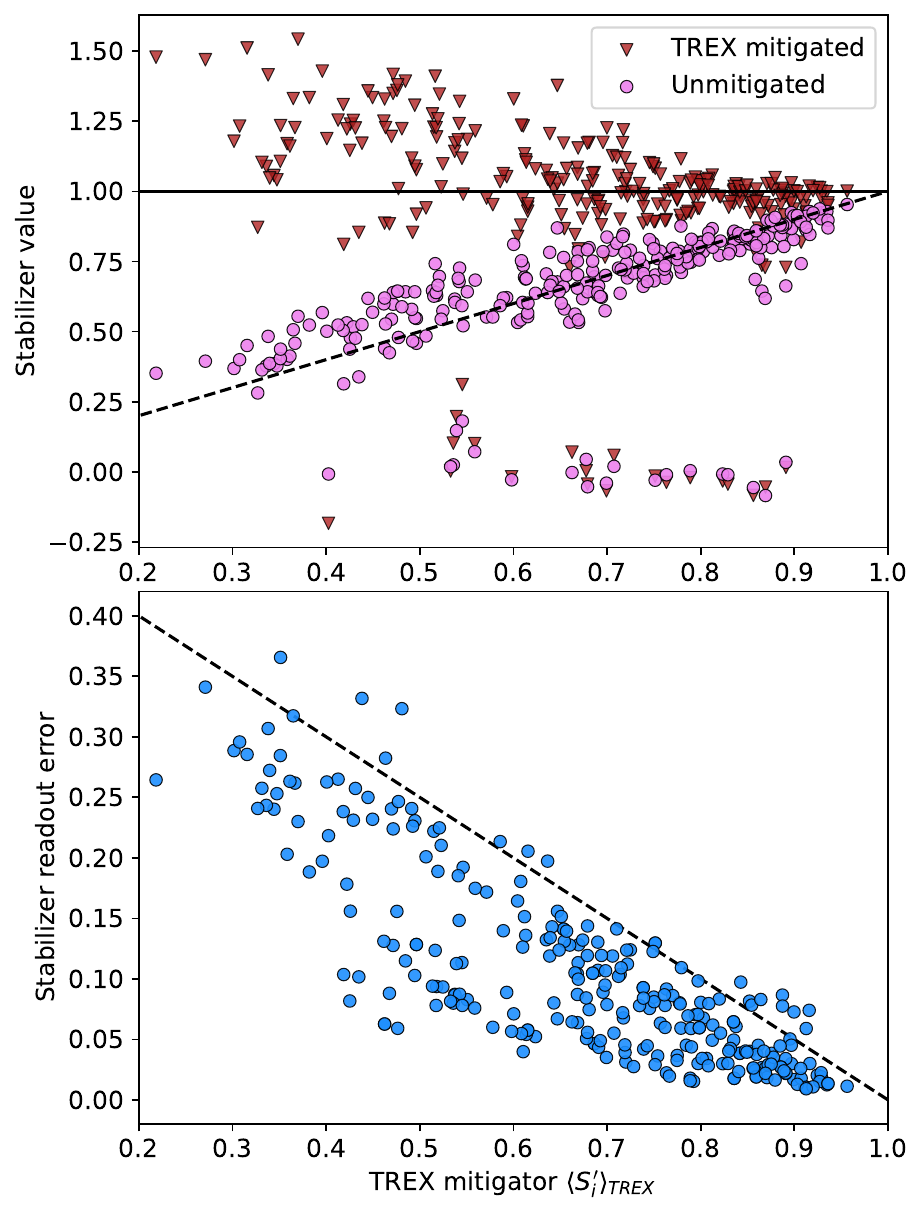}
    \caption{
    \textbf{Readout error mitigation with TREX.}
    (a)~The unmitigated stabilizer $\langle S_i\rangle_\text{Twirl}$ (pink circles) and the TREX mitigated stabilizers $\langle S_i\rangle_\text{Twirl}/\langle S'_i\rangle_\text{TREX}$ (red triangles) of the dropped edge data are plotted against the denominator $\langle S'_i\rangle_\text{TREX}$.
    Pink circles above the dashed line are stabilizers whose readout error is overestimated resulting in a mitigated stabilizer exceeding~1.
    The markers close to 0 correspond to stabilizers affected by a dropped edge.
    (b) Average stabilizer readout error $1-\prod_{j\in M_i}(1-r_j)$ as a function of $\langle S_i\rangle_\text{Twirl}$.
    Here, $M_i$ is the set of measured qubits contributing to stabilizer $S_i$ and $r_j$ is the readout error of qubit $j$.
    The dashed black line shows the expected relation between the readout error and $\langle S_i'\rangle_\text{TREX}$.
    }
    \label{fig:trex_effect}
\end{figure}

We observe that smaller $\langle S'\rangle_\text{TREX}$ denominators correlate with smaller unmitigated stabilizer values $\langle S\rangle_\text{Twirl}$, see Fig.~\ref{fig:trex_effect}(a).
This is expected as strong readout errors reduce the value of the observable and produce a smaller $\langle S'\rangle_\text{TREX}$, see Fig.~\ref{fig:trex_effect}(b).
However, TREX appears to overestimate the effect of strong readout errors resulting in denominators that are too small.
Indeed, the smaller the denominator the more likely it is that a TREX mitigated stabilizer will exceed 1, see red triangles in Fig.~\ref{fig:trex_effect}(a).
This can be due to state preparation errors arising, for instance, during qubit reset and $X$ gates.
TREX is benchmarked with $\mathcal{F}_\text{state}(P)\mathcal{F}_\text{meas.}(P)$ for different Pauli $Z$ terms $P$.
Here, $\mathcal{F}_\text{state}(P)$ and $\mathcal{F}_\text{meas.}(P)$ are the state preparation and measurement fidelity, respectively.
Under ideal state preparation, i.e., $\mathcal{F}_\text{state}(P)=1$, TREX isolates and corrects $\mathcal{F}_\text{meas.}(P)$.
However, when errors occur during state preparation, resulting in the preparation of the Pauli $Q$ instead of $P$, TREX will correct for $\mathcal{F}_\text{state}(Q)\mathcal{F}_\text{meas.}(Q)$.
This leads to an undesired factor of $\mathcal{F}_\text{state}(P)/\mathcal{F}_\text{state}(Q)$ which can be smaller than one ($Q$ has a smaller support than $P$) or larger than one ($Q$ has a larger support than $P$).
These effects can result in TREX-mitigated observable values that exceed one.

\begin{figure}
    \centering
    \includegraphics[width=\columnwidth]{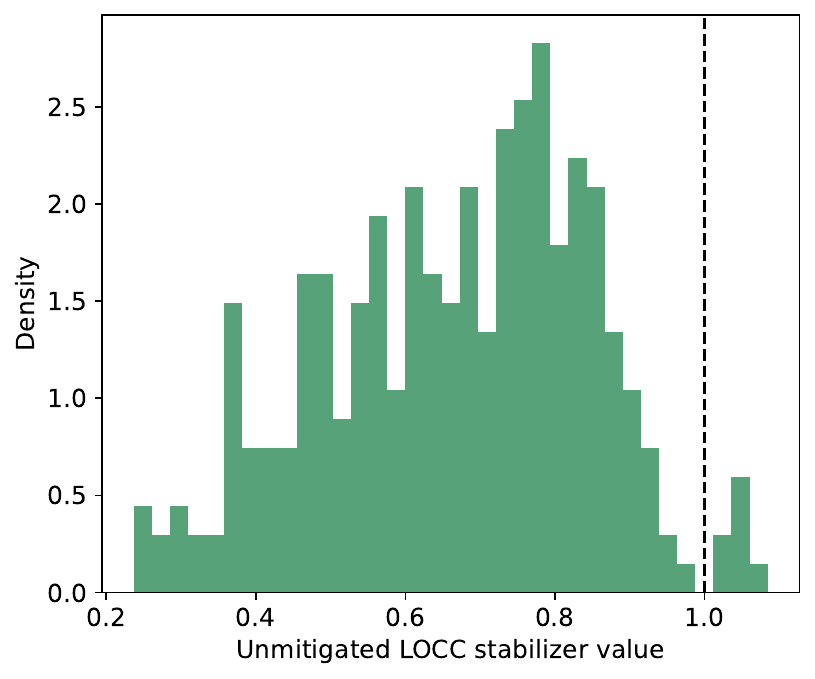}
    \caption{\textbf{LOCC stabilizers without error mitigation.}
    Distribution of the LOCC stabilizer values computed without the $\langle S'\rangle_\text{TREX}$ correction factor.
    The underlying data are obtained without stretching the delays in the switch, i.e., $c=1$.
    }
    \label{fig:qpd_effect}
\end{figure}

We now show that the QPD alone can result in stabilizer values that exceed 1 by analyzing the LOCC data of Fig.~\ref{fig:multi_qpu} of the main text without dividing by the TREX mitigator $\langle S'\rangle_\text{TREX}$.
Here, each channel $i=0,...,I-1$ in the QPD results in a count dictionary of bitstrings $\{b_j: f_j\}_i$ with $b_j\in\{0,1\}^n$ and $f_j\in\mathbb{N}$.
To reconstruct the value of the stabilizer we merge the $I$ count dictionaries by weighting the frequency of each bitstring by the QPD coefficients $a_i$.
This results in a single count dictionary
\begin{align}
    \left\{b_j: \sum_i a_i f_j\right\}
\end{align}
from which we compute the expectation value $\langle S_i\rangle$ of each stabilizer.
Here, since the QPD is not a valid probability distribution the resulting stabilizer can escape the $[-1, 1]$ interval.
For example, on the 134-node graph state we observe that 2.5\% of the 278 unmitigated LOCC stabilizers (measured at a switch stretch factor of $c=1$) have a value greater than 1, see Fig.~\ref{fig:qpd_effect}.
Since the stabilizers in this analysis are not normalized by $\langle S'\rangle_\text{TREX}$ we cannot attribute this deviation to TREX or the switch ZNE.
The lack of TREX correction also explains why the stabilizers in Fig.~\ref{fig:qpd_effect} have a large spread.

\section{Wall-clock time of LOCC vs LO}

To measure an observable $\langle O\rangle$ with a desired precision $\sigma_O$ we must gather enough shots to reach $\sigma_O$.
When cutting virtual gates, the $\gamma$ of the QPD is often used as a proxy for its cost with lower $\gamma$ being cheaper.
$\gamma$ is a reasonable cost proxy under the assumptions that (i) each circuit is assigned parameters, compiled, and then executed and that (ii) each circuit in the QPD has the same variance.
The situation changes when the quantum device can execute parametric payloads.
In this case, the backend receives a set of tasks in which task $k\in\{1, ..., K\}$ is composed of a single parameterized template circuit $C_k$ and an array of parameters with dimension $p\times S$.
Here, $p$ is the number of parameters in circuit $C_k$ and $S$ is the number of different sets of parameters to evaluate.
For simplicity, we assume here that the number of parameter sets $S$ is the same for each task $k$. 
This is the case for both LO and LOCC.
This execution model is favorable when all parameters are encoded in virtual $Z$-rotations since the template circuits $C_k$ are compiled once.
This work leverages two functions on the backend, \texttt{compile} and \texttt{execute}.
First, \texttt{compile} prepares the payload to be executed by \emph{compiling} a list of quantum circuits into a parameterized payload executable by the control electronics.
Crucially, the execution model accepts a multiplicity parameter $m$ such that on task $k$, the \texttt{compile} function compiles $m$ times the same template circuit $C_k$ into a single payload executable by the control electronics.
Next, \texttt{execute} is called to run $m$ of the $S$ parameter sets for which the desired number of shots are acquired.
As example, executing a single task with a circuit $\texttt{c1}$ and 20 parameter sets \texttt{p} and a multiplicity of $m=5$ proceeds as follows
\begin{lstlisting}
payload = compile([c1, c1, c1, c1, c1])
execute(payload, p[0:5])
execute(payload, p[5:10])
execute(payload, p[10:15])
execute(payload, p[15:])
\end{lstlisting}

\begin{figure}
    \centering
    \includegraphics[width=\columnwidth, clip, trim=0 0 30 20]{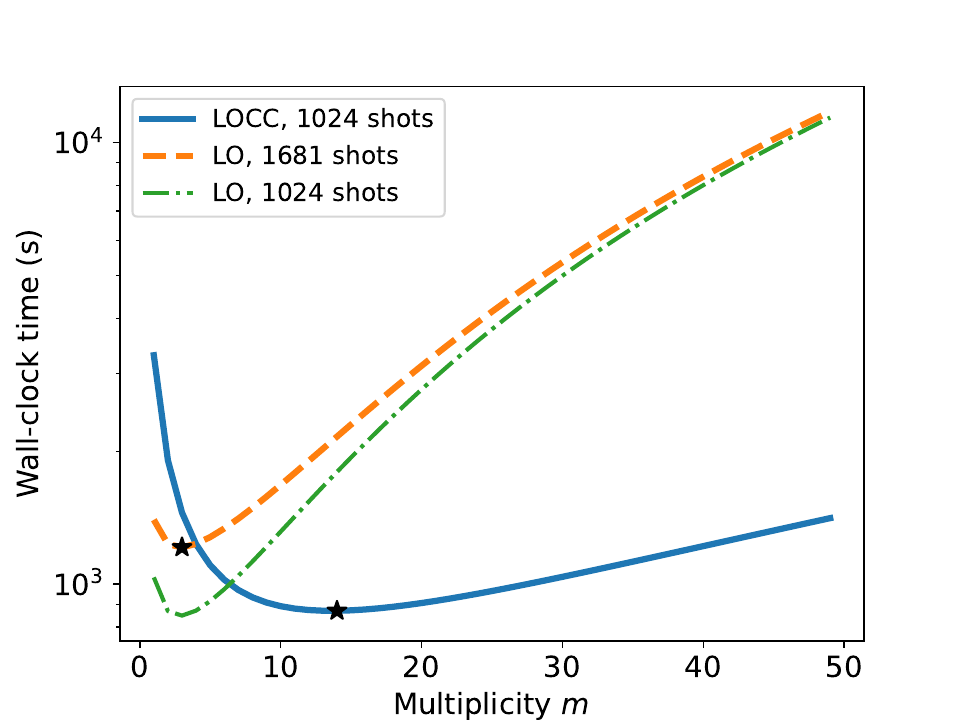}
    \caption{
    \textbf{Minimum wall-clock time of LO and LOCC}.
    The compile and execute time of LO and LOCC are measured and summed.
    The compile time shows a small quadratic dependence on $m$.
    The stars indicate the multiplicity that minimize the wall-clock time.
    The LO is displayed twice, once with the shot-overhead required to compensate for the $1.65\times$ larger $\gamma$ compared to LOCC (dashed orange) and once with the same number of shots per circuit as LOCC (dotted-dashed green).
    }
    \label{fig:wall_clock}
\end{figure}

We model the total wall-clock time of running LO and LOCC circuits as
\begin{align}
    t_\text{compile} + t_\text{execute} = f_c(m) + \frac{S}{m}(t_0 + m t_1).
\end{align}
Here, $f_c$ is a polynomial function modelling the compile time. 
The execute time is linear with parameters $t_0$ and $t_1$.
By properly choosing the multiplicity $m$ we can minimize this wall-clock time.
For different $m$, we measure the compile and execute time of circuits that create a 105-node graph state with periodic boundary conditions, similar to the one presented in the main text. 
We fit the compile and execute times to polynomial functions and find the multiplicity $m^*$ that minimizes their sum.
For LO and LOCC we obtain an $m^*$ of 3 and 14, respectively, see Fig.~\ref{fig:wall_clock}.
LO has a much lower $m^*$ than LOCC since it requires compiling nine different template circuits due to the presence or absence of the mid-circuit measurements shown in Fig.~\ref{fig:cz_lo}.
By contrast, LOCC requires compiling only a single template circuit (times the optimal multiplicity).

This benchmarking illustrates how significant the gains of a parametric execution are for large quantum circuits.
For instance, the LOCC circuit for a 105-node graph state compiles in $206~\mathrm{s}$ at the optimal multiplicity of $m^*=14$.
If each of the 4725 LOCC circuits were instead compiled at this $14.7~\mathrm{s/circuit}$ rate, then the total compile time is $19.30~\mathrm{hours}$. 
This highlights how crucial a parametric payload execution is since it reduces compile time down to $206~\mathrm{s}$ from $19.30~\mathrm{hours}$.

\section{Variance analysis\label{app:variance}}

\begin{figure}[h!]
    \centering
    \includegraphics[width=\columnwidth]{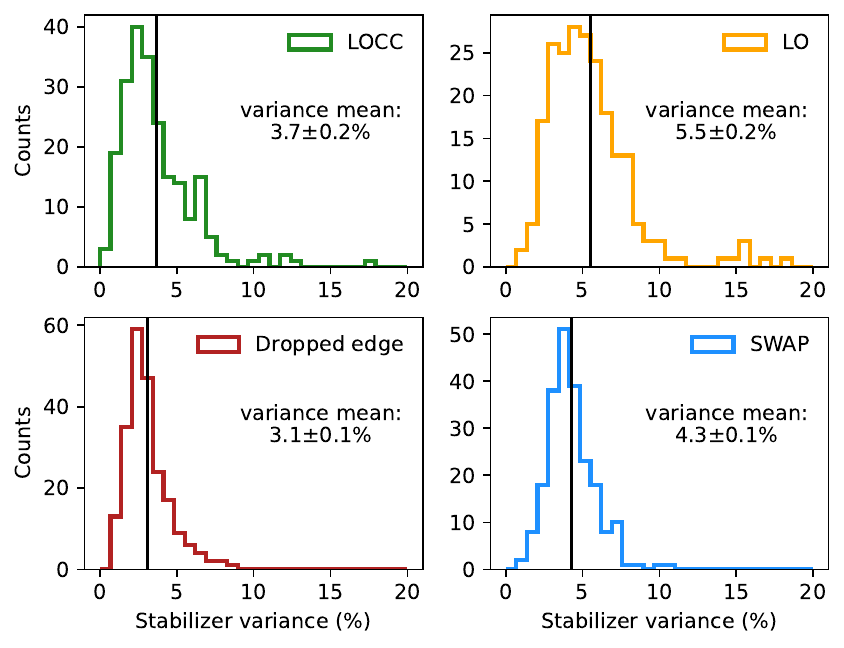}
    \caption{
    \textbf{Stabilizer variances on 103 nodes.}
    Distribution of the variance of the 216 measured stabilizers for the graph state with 103 nodes measured on \emph{ibm\_kyiv}.
    The black horizontal lines show the mean of the variance distribution with the errors corresponding to the standard error of the mean, i.e. $\sigma/\sqrt{N}$ with $N=216$.
    }
    \label{fig:stabilizer_variance}
\end{figure}

We now analyze the variance of the stabilizers measured with the dropped-edge benchmark, LO, and LOCC.
We observe that the stabilizers measured on graph states implemented with virtual gates have a variance that is comparable to those measured on graph states implemented with standard gates, see Fig.~\ref{fig:stabilizer_variance}.
For the 103-node graph, the average variance of a stabilizer measured with LOCC and LO is 3.7(2)\% and 5.5(2)\%, respectively.
The $\gamma$ of the LO and LOCC QPDs are 3 and 2.65 per cut gate.
Therefore, the relative variance increase of LO over LOCC for $n$ cut gates is $\gamma_\text{LO}^{2n}/\gamma_\text{LOCC}^{2n}$. 
In the main text we cut four gates with two QPDs executed in parallel.
We therefore expect the average variance of a stabilizer measured with LO to be $9^2/7^2=1.65\times$ the average variance of a stabilizer measured with LOCC since the circuits in both methods are executed with the same number of shots, i.e., 1024.
Based on the measured LO and LOCC variances the measured increase is $1.49(8)\times$.
This is slightly smaller than the $1.65\times$ expected increase.
We attribute this discrepancy to the additional noise present in the dynamic circuits in LOCC caused by the extra latency.
Nevertheless, the measured value is in good agreement with the theoretically expected value and justifies current efforts on finding QPDs with a low $\gamma$-cost.
However, we caution that the LO and LOCC data were acquired on different days during which the noise of \emph{ibm\_kyiv} may have drifted.

\begin{figure*}
    \centering
    \includegraphics[width=\textwidth]{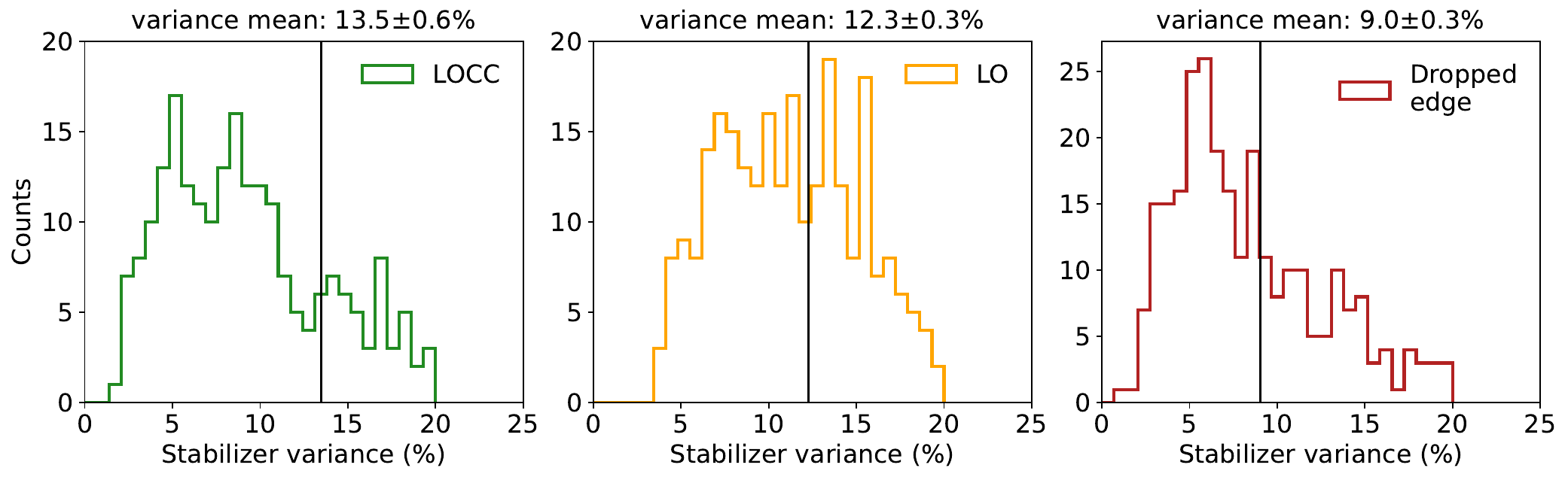}
    \caption{
    \textbf{Stabilizer variances on 134 nodes.}
    Distribution of the variance of the 278 measured stabilizers for the graph state with 134 nodes measured on \emph{ibm\_pinguino\_2a}.
    The black horizontal lines show the mean of the variance distribution with the errors corresponding to the standard error of the mean, i.e. $\sigma/\sqrt{N}$ with $N=278$.
    }
    \label{fig:stabilizer_variance_p23}
\end{figure*}

We also analyze the variances of the measured stabilizers of the 134-node graph acquired on $\emph{ibm\_pinguino-2a}$.
The variances obtained with LO are lower than those obtained from LOCC, see Fig.~\ref{fig:stabilizer_variance_p23}.
In addition, the stabilizers measured with LO have smaller errors than those in the dropped-edge benchmark despite the mid-circuit measurements of the LO circuits, see Fig.~\ref{fig:multi_qpu} in the main text.
We attribute this better-than expected performance of LO to experimental drifts.
When comparing the average variance increase of LO over LOCC we obtain a factor of $0.91(6)$ which is in disagreement with the expected $1.65\times$ increase.
We attribute this deviation to the fact that the LO and LOCC data sets were gathered 22 days apart during which the device noise may have changed.
Future work should thus perform a careful systematic variance characterization and analysis to compare the true experimental cost of LO relative to LOCC.
This is a complex undertaking due to the different noise sources in both protocols.

\clearpage

\bibliography{references}

\end{document}